\documentclass[sigconf,nonacm]{acmart}

\AtBeginDocument{%
  }

\acmDOI{} \acmISBN{} \acmBooktitle{} \acmPrice{}  
\acmConference[]{}{}{}

\usepackage{ifthen}
\usepackage{xspace}
\usepackage{subcaption}
\usepackage[normalem]{ulem} 
\usepackage[nolist,nohyperlinks]{acronym}
\usepackage{soul}
\usepackage{listings}

\usepackage{color}
\newboolean{authorcomments}
\setboolean{authorcomments}{true}
\newcommand{\bssm}[1]{\ifthenelse{\boolean{authorcomments}}{{\color{blue}\textit{[#1 -ssm-]}}\xspace}{}}
\newcommand{\rssm}[1]{\ifthenelse{\boolean{authorcomments}}{{\color{red}\textit{[#1 -ssm-]}}\xspace}{}}
\newcommand{\gssm}[1]{\ifthenelse{\boolean{authorcomments}}{{\color[rgb]{0,0.6,0}\textit{[#1 -ssm-]}}\xspace}{}}

\newboolean{long} 
\newcommand{\lvsv}[2]{\ifthenelse{\boolean{long}}{#1}{#2}} 
\newboolean{anonymous} 
\setboolean{anonymous}{true}
\newcommand{\onlyan}[1]{\ifthenelse{\boolean{anonymous}}{#1}{}} 
\newcommand{\onlynonan}[1]{\ifthenelse{\boolean{anonymous}}{}{#1}} 
\newcommand{\annonan}[2]{\ifthenelse{\boolean{anonymous}}{#1}{#2}} 

\begin{document}

\title{AI-Assisted Modeling: DSL-Driven AI Interactions}

\author{Steven Smyth, Daniel Busch, Moez Ben Haj Hmida, Edward A. Lee, Bernhard Steffen}

\settopmatter{printacmref=false}
\renewcommand\footnotetextcopyrightpermission[1]{}

\email{steven.smyth@cs.tu-dortmund.de, daniel.busch@cs.tu-dortmund.de}
\email{moez.benhajhmida@enit.utm.tn, eal@berkeley.edu, bernhard.steffen@cs.tu-dortmund.de}

\renewcommand{\shortauthors}{Smyth et al.}

\acrodef{ai}[AI]{Artificial Intelligence} \acused{ai}
\acrodef{api}[API]{Advanced Programming Interface} \acused{api}
\acrodef{dsl}[DSL]{Domain-specific Language} \acused{dsl}
\acrodef{hal}[HAL]{Holistic Abstraction Layer}
\acrodef{http}[HTTP]{Hypertext Transfer Protocol} \acused{http}
\acrodef{ide}[IDE]{Integrated Development Environment} \acused{ide}
\acrodef{ip}[IP]{interaction point} 
\acrodef{lde}[LDE]{language driven engineering} 
\acrodef{llm}[LLM]{Large Language Model} \acused{llm}
\acrodef{moc}[MoC]{Model of Computation}
\acrodef{oop}[OOP]{Object-oriented Programming} \acused{oop}
\acrodef{op}[OP]{observation point} 
\acrodef{php}[PHP]{PHP Hypertext Preprocessor} \acused{php}
\acrodef{rag}[RAG]{Retrieval-Augmented Generation}
\acrodef{sql}[SQL]{Simple Query Language} \acused{sql}
\acrodef{uml}[UML]{Unified Modeling Language} \acused{uml}

\begin{abstract}
AI-assisted programming greatly increases software development performance. We enhance this potential by integrating transparency through domain-specific modeling techniques and providing instantaneous, graphical visualizations that accurately represent the semantics of AI-generated code. This approach facilitates visual inspection and formal verification, such as model checking.

Formal models can be developed using programming, natural language prompts, voice commands, and stage-wise refinement, with immediate feedback after each transformation step. This support can be tailored to specific domains or intended purposes, improving both code generation and subsequent validation processes.

To demonstrate the effectiveness of this approach, we have developed a prototype as a Visual Studio Code extension for the Lingua Franca language. This prototype showcases the potential for novel domain-specific modeling practices, offering an advancement in how models are created, visualized, and verified.
\end{abstract}



\maketitle

\section{Introduction}
\label{sec:introduction}

Similar to general \ac{ai}-assisted programming---which primarily targets classical, textual programming languages and is exemplified by tools such as Copilot\footnote{\url{https://github.com/features/copilot}} and Cursor\footnote{\url{https://www.cursor.com}}---domain-specific modeling languages are beginning to benefit from growing \ac{ai} support through natural language extensions.
Recent work by Busch et al.~\cite{BuschNBS23, BuschBSS24} demonstrates such an extension by combining informal, intuitive natural language with formally defined graphical domain-specific languages (DSLs). 
This is achieved by adapting the code generator of the modeling language to produce structured prompt templates that allow \acp{llm} to generate implementation code for dedicated purposes that is automatically integrated in the overall system. Validation is realized at the system level via automata learning~\cite{SteffenHM11}, which ensures the generated, global models align with expected behavior.

In this paper, we enhance AI-assisted programming with domain-specific modeling techniques and providing instantaneous, graphical visualizations that accurately represent the semantics of AI-generated code as a basis for visual inspection and formal verification, such as model checking.
Our approach to AI-assisted programming is hybrid:

\noindent\mbox{ -- }Technically, it builds on state-of-the-art, AI-assisted, integrated development environments, similar to the agents one sees in \acp{ide} like VS Code and Cursor. 
The adoption of these agents has drastically changed (and in the best case increased) the development performance of experts by automatically generating and adapting code in response to natural language descriptions. 
This includes the search for adequate available functionality, provides explanations of certain phenomena, and inserting documentation.

\noindent\mbox{ -- }Conceptually, we leverage domain-specificity to enhance control. 
This is particularly powerful when combined with service-orientation, where complex features are provided in a well-under\-stood form as programming or modeling components. 
Combined with hierarchy this enables modularity-based scalability and validation.

\noindent\mbox{ -- }Pragmatically, we support instantaneous, graphical visualizations where developers can continuously follow the impact of their design steps by inspecting automatically generated graphical models that are semantically accurate. 
Besides this accurate visualization, which typically concerns the unit level, we also provide system-level hypothesis models on the basis of automata learning~(cf. \cite{BuschBSS24}).

We demonstrate the approach through a working \ac{ide} extension for Visual Studio Code and compatible \acp{ide}, applied to the Lingua Franca programming language for cyber-physical systems.
Lingua Franca's polyglot nature presents particular challenges for \acp{llm}, making it a compelling case for incremental, context-aware modeling rather than single-shot generation.
In the extension, formal models can be developed using \ac{dsl} programming, natural language prompts, and voice commands, which form stage-wise refinements with immediate feedback after each transformation step. 
This combines domain-specific programming and modeling with graphical visualization and validation.

\paragraph{Background}
While \ac{ai}-assisted programming offers powerful new capabilities and continues to deliver impressive results, it remains difficult to control and steer effectively.
With increasing complexity, developers face growing challenges in maintaining oversight and responding in a timely and accurate manner---especially when relying on \emph{one-shot prompts}.
Moreover, classical modeling workflows frequently depend on graphical representations, which are not easily generated by current \ac{ai} systems.

Although generating entire models from generic prompts can be powerful, \ac{ai} assistants often fail to produce correct results on the first attempt.
This is particularly true for \acp{dsl} with uncommon syntax or domain-specific concepts that are underrepresented in training data.
For instance, the polyglot nature of Lingua Franca~\cite{LohstrohMBL21} and its use of host code delimiters (\verb|{= =}|) pose significant challenges for general-purpose \acp{llm}.
Even in straight-forward \acp{dsl}, structural peculiarities---such as transitions defined outside a superstate’s scope in the SCCharts language~\cite{vonHanxledenDM+14}---further complicate automated generation and step-wise refinement.

\begin{figure}[tb]
	\includegraphics[width=1\linewidth]{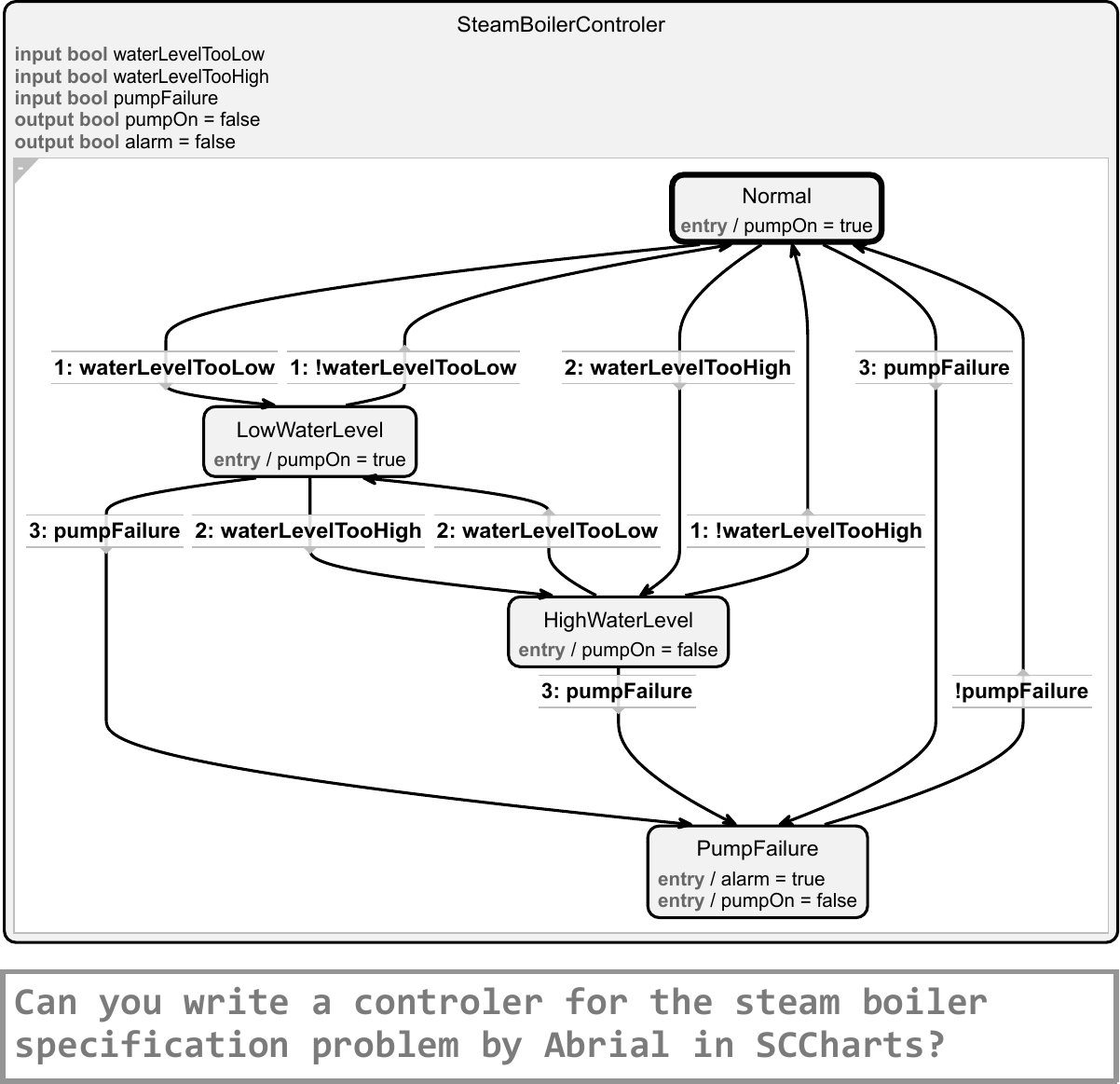}
	\caption{SCCharts one-shot \ac{ai} Steamboiler with prompt (Smyth, \textsc{\smaller{SYNCHRON'23}}) referring to the specification problem from Abrial~\cite{Abrial05}}
	\label{fig:sccharts-steamboiler}
\end{figure} 

\begin{figure*}[tb]
	\includegraphics[width=1\linewidth]{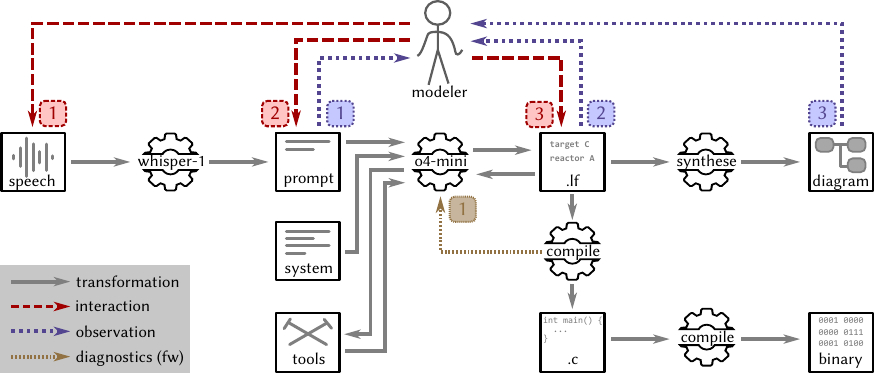}
	\caption{The modeling concept of \ac{dsl}-driven \ac{ai} interaction: A modeler interacts with the system through speech or text inputs and receives feedback from multiple transformation stages, including instantaneous diagram synthesis and compilation results. This workflow forms a feedback loop, enabling both the modeler and the system to incrementally refine the model.}
	\label{fig:user-story}
\end{figure*}

\paragraph{Solution}
We combine (i) state-of-the-art \ac{dsl} modeling techniques with instantaneous automatic visualizations, and (ii)~\ac{ai}-assisted programming to enable iterative and interactive support for \ac{dsl} development.

(i) Textual domain-specific (natural) languages are well suited to guide AI assistants---much like AI-assisted pair-programming in general-purpose languages.
In contrast, generating graphical models via \ac{ai} is more technically demanding, though such representations offer immediate overviews, intuitive navigation, and visual diagnostics that can help to identify errors quickly.
Such graphical representations that correspond to textual models can be synthesized automatically.
For example, using a prompting technique similar to that described by Busch et al., the SCCharts language~\cite{vonHanxledenDM+14} for safety-critical systems can be leveraged to generate complete textual models, which are then synthesized into diagrams automatically, as shown in~\autoref{fig:sccharts-steamboiler}.
In this example, an \ac{llm} is prompted to implement a steam boiler controller based on Abrial's specification~\cite{Abrial05} without any further guidance.
While the result is not complete, it serves as a valuable starting point for the modeler.

(ii) To address the growing complexity of \ac{ai}-assisted \ac{dsl} modeling, we propose increasing the number of \emph{observation} and \emph{interaction points} throughout the modeling process.
These points go beyond a dedicated single chat window to interact with the agent as it is currently the state-of-the-art. 
They allow modelers to inspect intermediate results, guide refinements, and provide feedback---supporting more flexible, iterative workflows that move beyond rigid one-shot prompting.
Inspired by the divide-and-conquer approach in classical \ac{lde}, we decompose the program/model refinement process into distinct stages, allowing for incremental progress and improved controllability during model development.

\paragraph{Contribution}
We present an approach for instantaneous, visual verification tailored to \ac{dsl} development, leveraging observation points across modeling stages and enabling stepwise program/model refinement.
This supports a fine-grained, interactive workflow that improves the reliability and controllability of model generation.

We demonstrate the approach through a working \ac{ide} extension for Visual Studio Code and compatible \acp{ide}, applied to the Lingua Franca programming language for cyber-physical systems.
Lingua Franca's polyglot nature presents particular challenges for \acp{llm}, making it a compelling case for incremental, context-aware modeling rather than single-shot generation.

Finally, we outline how parts of this extension may themselves be generated via AI in the future, further streamlining the development process.

\paragraph{Outline}
\autoref{sec:future-of-modeling} presents our approach to AI-assisted modeling, outlining the overall process and highlighting the role of observation and interaction points in enabling iterative refinement.
The section concludes with a discussion of key implications for modeling workflows.
Implementation details and the Visual Studio Code demonstrator are provided in \autoref{sec:implementation}.
Related work is reviewed in \autoref{sec:related-work}, and we conclude in \autoref{sec:conclusion}.
Complementing this paper, a demonstration video is available online\footnote{\url{https://youtu.be/dQVFTaYciUQ}}.

\setul{0.5ex}{0.3ex}
\newcommand{\incolor}[2]{{\color{#1}#2}}
\newcommand{\oncolor}[2]{\setulcolor{#1}\ul{#2}}
\newcommand{\tsf}[1]{\textsf{#1}}
\newcommand{\ip}[1]{\raisebox{-0.2ex}{\includegraphics[scale=0.7]{images/#1.png}}}

\newcommand{\req}[1]{\textbf{(R#1)}}

\section{A Modeling Future}
\label{sec:ai-assisted-modeling}
\label{sec:future-of-modeling}

AI-assisted programming has rapidly evolved from simple question-answer interactions into rich, dialog-based pair-programming experiences with \acp{llm}.
Building on these developments, this work investigates how similar forms of assistance can be applied to domain-specific modeling, with the goal of enabling a more natural and interactive modeling experience.
Instead of treating model generation as a one-shot task, our approach introduces a dialog-driven workflow that facilitates clarification, program/model refinement, and visual feedback throughout the process.

Modern state-of-the-art tools, such as Cursor and other specifica\-tion-driven approaches, already follow an iterative approach and move away from one-shots. 
Commonly, they formulate a plan first and the agent then attempts to solve the given prompt though multiple iterations relying on the diagnostics of the \ac{ide}. 
The developer then can refine the results. 
We aim to involve the developer even more deeply in the iterations and refine individual steps early on.

\paragraph{The Concept}
The overall concept is illustrated in \autoref{fig:user-story}.
The \tsf{modeler}, who interacts with the system through speech or text inputs and receives feedback from multiple \tsf{transformation} stages, is the central figure in this process.
A typical transformation workflow begins at the \tsf{speech} input stage, where recorded speech is transcribed and interpreted by an \ac{llm} into a structured \tsf{prompt} frame.
Our preliminary tests~(cf.~\textsuperscript{3}) suggest that the reaction is already reasonably fast and offers new modeling opportunities---particularly for non-experts.
The prompt may be refined by the developer before sending the text to the \ac{llm} or send immediately for \emph{very rapid prototyping}, in which case the speech basically becomes the sole input for the next program/model refinement step. 
This also enables additional capabilities, such as automatic translation, which would enable 'programming' in ones native language.
Given that many \acp{llm} are multilingual, this process should be relatively seamless, assuming the transcription model supports your language accurately.
 
At the core of the process workflow, the \tsf{user prompt}, the pre-defined (Lingua Franca) \tsf{system prompt}, including the \tsf{tools} definitions (via the OpenAI\footnote{https://platform.openai.com} Tool \ac{api}), and the current (\tsf{.lf}) model are passed to the \ac{llm}.
This can be extended by any additional information of the domain, further workspace files, and also chat history if testing indicates better results.

\begin{figure*}[tb]
	\includegraphics[width=1\linewidth]{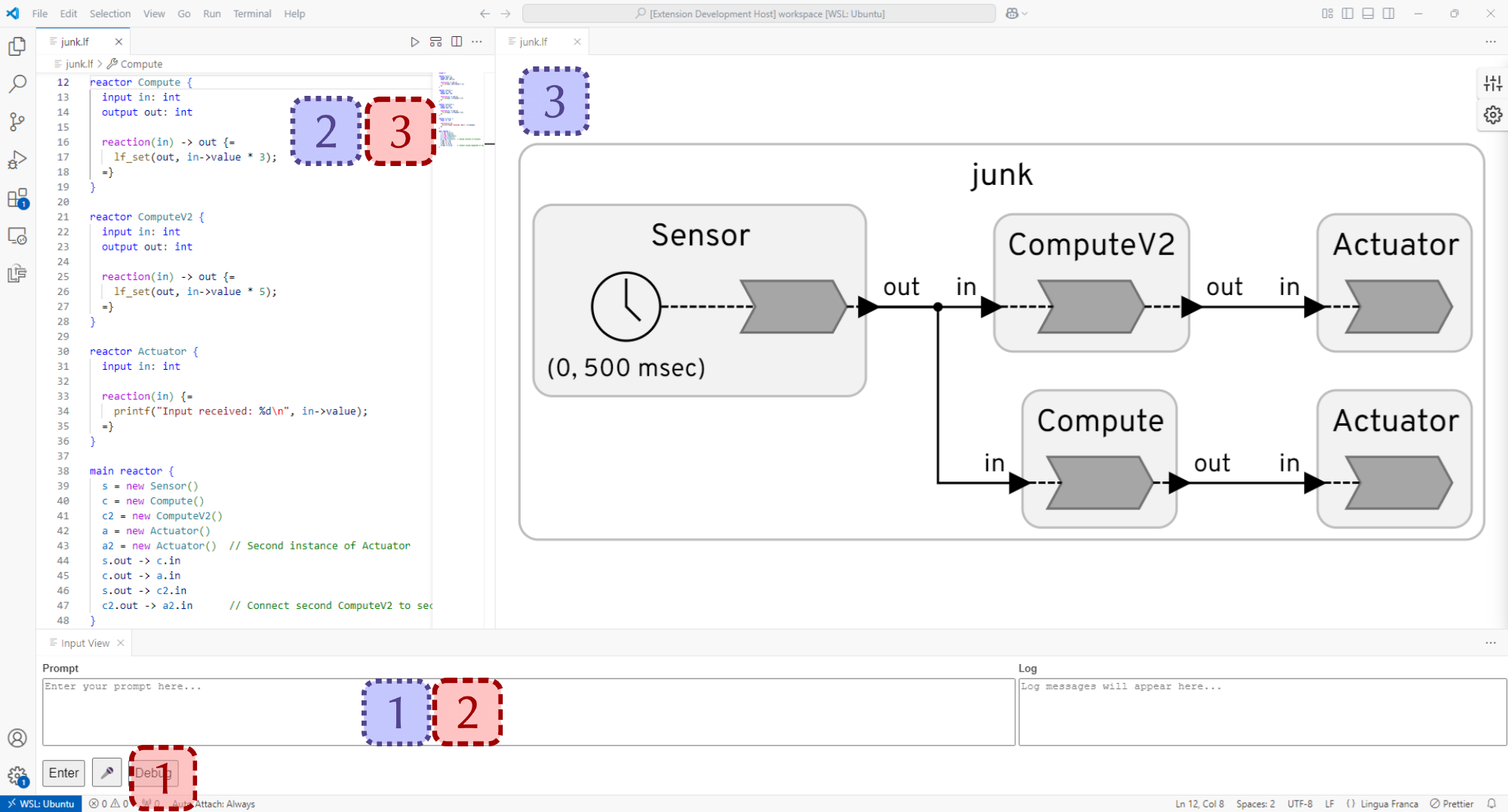}
	\caption{Demonstrator of rapid, natural modeling with Lingua Franca in Visual Studio Code. The observation points~\ip{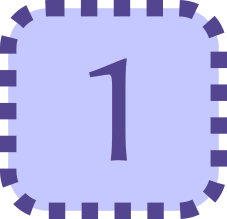}, \ip{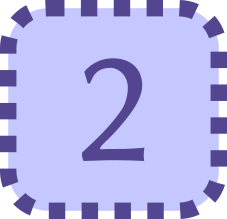}, and \ip{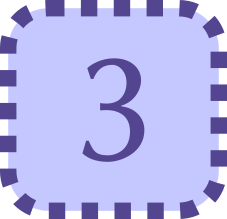} display the outputs of the speech transcription, the generated textual model, and the synthesized diagram, respectively. The interaction points~\ip{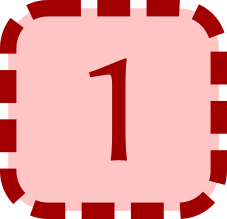}, \ip{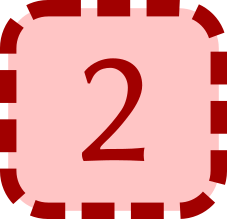}, and \ip{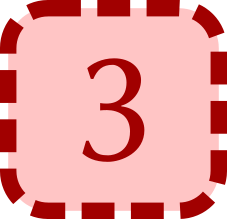} allow the modeler to provide input via speech, text prompt, or direct editing of the model.}
	\label{fig:demonstrator}
\end{figure*}

One concrete way to retrieve correct syntax from the LLM is by leveraging the Tool \ac{api}~\cite{BuschNBS23}.
Rather than encoding extensive context into prompts to convey the semantics and capabilities of a \ac{dsl}, we shift the burden of \emph{contextualization} to the tool itself.
Through the \ac{api}, an \ac{llm} can perform predefined actions and receive concrete feedback, such as querying internal state, generating precise syntax, or applying transformations.
This is particularly useful for \acp{dsl} with uncommon syntax or polyglot constructs.
The prototype shown here solely relies on the static system prompt and the tools \ac{api} that provides concrete source code for single language elements, which can be generated from the language grammar (see \autoref{sec:implementation}).
Note, however, that leveraging the Tool \ac{api} is not mandatory and other retrieval techniques, such as \ac{rag} or \ac{llm} fine-tuning, or a combination of these can also lead to better results.
Determining the optimal granularity of Tool \ac{api} functions and combination of techniques to retrieve reliable, scalable syntax remains an open question; \autoref{fig:demonstrator} illustrates a case where the API's granularity is aligned with the grammar rules of the \ac{dsl}.
It is worth noting that defining ``enough context'' for classical prompt engineering presents a similarly difficult challenge.

The \ac{llm} then produces a new textual working model, which is presented as usual in the \acp{ide} editor. 
It can be modified and serves as artefact for typical development pipelines.
Hence, the model can serve as source of a code generator to produce executable code, which may then be used for deployment, simulation, or debugging, depending on the capabilities of the backend.
Simultaneously, it is automatically and instantaneously synthesized into a graphical \tsf{diagram} to provide instant visual feedback to help the developer to gain an overviews and enable targeted refinements without the need to dig through the whole document.

This workflow naturally forms a feedback loop, enabling both the modeler and the system to incrementally refine the model.
The effectiveness of these refinements depends on the availability of observation and interaction points throughout the process.

\paragraph{Observation and Interaction}
Several stages of the aforementioned workflow offer natural opportunities for observation and interaction. The list can be extended, e.g., with intermediate results from the chat history or diagnostics, in the future.

The \tsf{speech} \ac{ip}~\ip{r1} serves as the entry point, allowing the modeler to interact with the system conversationally---facilitating rapid prototyping in a natural way.
The transcription result is displayed in the prompt frame~\ip{b1}, the first \ac{op}.
If the transcription is unsatisfactory, the modeler can retry the speech input via \ip{r1} or proceed with editing the prompt at \ip{r2}.

Once the prompt is confirmed, the system uses all available inputs---including the prompt, the \ac{dsl}'s system and tool definitions, and the current model state---to generate a new working model.
This intermediate result is displayed in the standard text editor of the \ac{ide}~\ip{b2}, where typical features such as editing, syntax highlighting, and validation, are available~\ip{r3}.

For \acp{dsl} with graphical representations, the textual model is synthesized into a diagram using transient views and automatic layout techniques~\cite{SchneiderSvH13}.
This graphical view~\ip{b3} provides immediate, comprehensible feedback and supports intuitive navigation.

Compilation results act as an additional form of visual verification and therefore represent another \ac{op}.
These outputs—such as compiler diagnostics and analysis results—can inform subsequent refinement steps~\ip{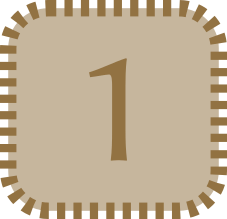}, enabling more targeted assistance by the \ac{llm}.
This type of feedback loop has already been demonstrated in the Cursor \ac{ide} and others, where compiler messages are integrated into the assistant's reasoning process.
Note that this observation point is not yet utilized in the current prototype; the model refinement is currently based solely on the model state, the modeler's input, and the pre-defined language semantics.

Note that the presented list of observation and interaction points is not meant to be complete, but rather provides a reasonable collection for textual \acp{dsl} to include the developer more closely into the iterative program/model refinement process.
Depending on the domain and the available tooling, this list should be extended to increase the granularity.
Furthermore, the \ac{llm} chat history can be enriched with the results of all intermediate steps---textually and graphically---providing further documentation of the whole development process.

\paragraph{Discussion}
The main advantage of this approach lies in its support for iterative program/model refinement and rapid correction.\footnote{This is also demonstrated in the first accompanying video\textsuperscript{3}, where the assistant does not automatically set the mandatory target language in Lingua Franca, but the modeler can immediately respond.}
Unlike black-box, one-shot prompt engineering, our method allows the modeler to inspect individual stages of the transformation process and, ideally, understand the reasons behind intermediate results.
Each corrective action helps guide the assistant toward more accurate refinements in subsequent iterations and documents the progress.

All observation points are available simultaneously, and interactions can occur at the corresponding interaction points where needed.
In the future, this will also be extended to the whole chat history of the developer, which provides the \ac{llm} with a detailed description of the development progress.
Transformation stages do not need to be manually triggered for each refinement cycle.
In a practical example,\textsuperscript{3} most interactions take place at the entry point~\ip{r1}, which then automatically run through all phases and the result is typically observed in the diagram~\ip{b3}.
The modeler can, however, also interact with any intermediate stage when needed.

\begin{figure*}[tb]
	\includegraphics[width=1\linewidth]{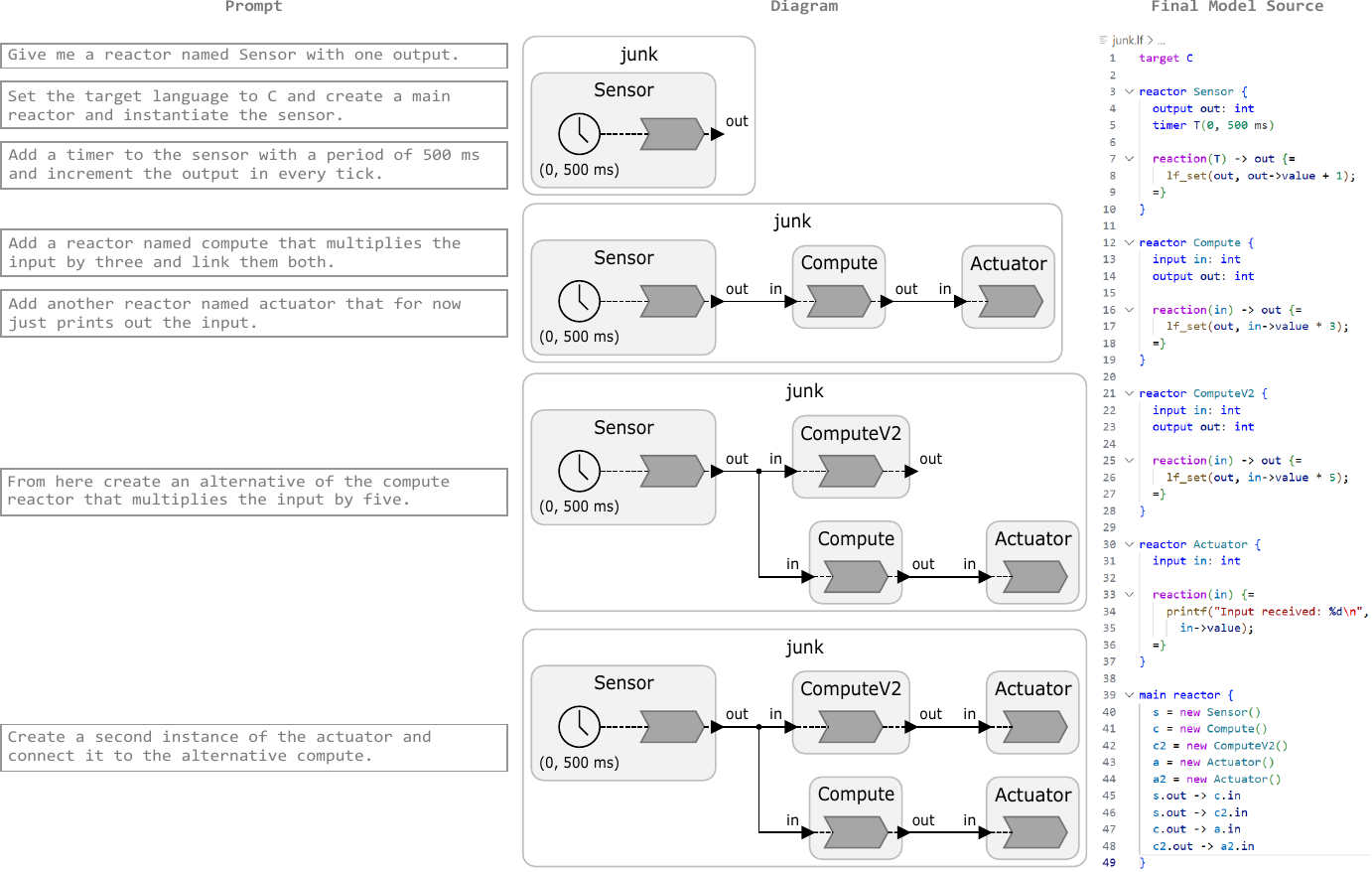}
	\caption{Demonstration of the prototype in action, following the transcript from the accompanying video\textsuperscript{3}. The time progresses from top to bottom and the model was created in less than 2 min. All prompts on the left side were spoken by the modeler, transcribed, and then sent to the \ac{llm} to generate the Lingua Franca source model, which is then synthesized into the corresponding diagram. The final Lingua Franca model source is shown to the right.}
	\label{fig:demo-transcript}
\end{figure*}

Graphical representations are particularly effective at conveying complex relationships, which makes them especially useful in \ac{dsl} design.
This approach combines the clarity of diagrams with the generative advantages of textual \acp{dsl}, which are well-suited for \ac{llm}-based synthesis.
We further argue that natural speech input and comprehensible, easy-to-navigate diagrams are essential for enabling a fast and accessible modeling experience.
To strengthen a pair-modeling experience even more, further speech commands for navigation and modification can be implemented in the future.

Finally, the general output of the \ac{llm} can further be improved by fine-tuning- and/or retrieval-based techniques. 
These are orthogonal to the modeling methodology presented in this work and current research in these areas will likely contribute to the overall modeling experience.

\section{Prototype Demonstrator}
\label{sec:implementation}
\label{sec:demonstrator}

Our initial prototype is realized as Visual Studio Code extension and follows the concept, outlined in~\autoref{sec:future-of-modeling} straightforwardly. Yet, it already suggests further opportunities for improvement---particularly at the meta level, which are discussed at the end of the section.

\autoref{fig:demonstrator} shows a running instance of the editor with activated extension. 
In addition to the standard \ac{ide} elements, such as the editor and the toolbar to the left, a webview containing the synthesized diagram to the right and a view showing the \ac{ai} interaction at the bottom are visible.
The observation and interaction points introduced in \autoref{sec:future-of-modeling} are annotated in the figure.
They are designed to be intuitive and to integrate easily in common workflows.
The modeler can either use the voice input~\ip{r1}, the natural language prompt~\ip{r2}, or the text editor~\ip{r3} to input their program/model refinements. 
Accordingly, the outputs of the \acp{llm} appear in the prompt~\ip{b1}, the text editor~\ip{b2}, and the instantaneously generated diagram~\ip{b3}.

\autoref{fig:demo-transcript} illustrates the demonstrator in action, following the transcript from the accompanying video\textsuperscript{3} and showcasing the seamless integration of speech, text, and visual modeling. 
The time progresses from top to bottom and the model was created in less than 2 min.
All prompts on the left side were spoken by the modeler, transcribed, and then sent to the \ac{llm} to generate the Lingua Franca source model, which is then synthesized into the corresponding diagram. 
A diagram is generated from the prompts next to it. 
The final Lingua Franca model source is shown to the right.

\paragraph{Implementation}
As illustrated in \autoref{fig:user-story}, the prototype uses the \tsf{whisper-1} model for the speech-to-text transcription and \tsf{o4-mini} for the model creation. 
These models are interchangeable and can be replaced with alternatives if desired.
The Lingua Franca diagram is synthesized automatically via the Lingua Franca extension, which relies on the KIELER diagram engine, built on the Eclipse Layout Kernel (ELK).\footnote{https://eclipse.dev/elk}

Preliminary tests, as demonstrated in the accompanying video\textsuperscript{3}, suggest that the workflow is fluent and easy to follow.
The speech-to-text transcription is almost instant and offers a novel way to interact with the modeling environment and enables, in particular, non-experts to quickly prototype models.
This may also provide an additional correction layer for non-native speakers.
For the model creation, currently the \tsf{o4-mini} model is used to shorten response time. 
Although tool calls introduce some overhead, the improved quality of the results justifies the trade-off---particularly since generating a correct model suitable for diagram synthesis is the primary objective.

Please note that the use of specific \ac{api} vendors and models is not decisive for the methodology.
The approach can be implemented via common \ac{ai} \acp{api}, such as Vercel's\footnote{https://ai-sdk.dev}, and comparable \acp{llm} depending on your \ac{api} access.

\paragraph{Future Meta Tooling}

\begin{lstlisting}[float, floatplacement=b, basicstyle=\scriptsize\ttfamily, caption={Manually written Tool \ac{api} description for a Lingua Franca timer},label={lst:lf-timer-tool-api}]
// timer ${args.name}(${args.offset}, ${args.period});
{
	type: 'function',
	function: {
		name: 'createTimer',
		description:
		'Returns the code for a timer with the provided name, offset, and period. Unless 0, a unit must be provided. Typical units are s, ms, us, and ns.',
		parameters: {
			type: 'object',
			properties: {
				name: {
					type: 'string',
					description: 'Name of the timer. Must be unique within the reactor. The name is used to refer to the timer in the reaction. Typical name is T.',
				},
				offset: {
					type: 'string',
					description: 'The offset of the timer as string because a unit must be provided unless value is 0.',
				},
				period: {
					type: 'string',
					description: 'The period of the timer as string because a unit must be provided unless value is 0.',
				},
			},
		},
	},
}
\end{lstlisting}

\begin{lstlisting}[float, floatplacement=b, basicstyle=\scriptsize\ttfamily, caption={Current timer definintion in the Lingua Franca grammar},label={lst:lf-timer-grammar}]
// Timing specification for a timer: (offset, period)
// Can be empty, which means (0,0) = (NOW, ONCE).
// E.g. (0) or (NOW) or (NOW, ONCE) or (100, 1000)
// The latter means fire with period 1000, offset 100.
Timer:
	(attributes+=Attribute)* 'timer' name=ID ('(' offset=Expression (',' period=Expression)? ')')? ';'?;
\end{lstlisting}

\begin{lstlisting}[float, floatplacement=b, basicstyle=\scriptsize\ttfamily, caption={Tool \ac{api} description for a Lingua Franca timer generated from comments in the grammar file},label={lst:lf-timer-tool-api-generated}]
	{
		"name": "createTimer",
		"description": "Creates a timer definition in concrete syntax based on the given parameters.",
		"parameters": {
			"name": {
				"type": "string",
				"description": "The name of the timer."
			},
			"offset": {
				"type": "integer",
				"description": "The offset before the first trigger. Default is 0 (NOW).",
				"optional": true
			},
			"period": {
				"type": "integer",
				"description": "The period between subsequent triggers. Default is 0 (ONCE).",
				"optional": true
			},
			"attributes": {
				"type": "array",
				"items": {
					"type": "string"
				},
				"description": "Optional attributes for the timer.",
				"optional": true
			}
		},
		"returns": {
			"type": "string",
			"description": "The concrete syntax representation of the timer."
		}
	}
\end{lstlisting}

While the concrete syntax returned by a Tool \ac{api} call may consist of a single line, the accompanying description for the \ac{llm} is often considerably more extensive.
\autoref{lst:lf-timer-tool-api} shows the description to create a timer in Lingua Franca.
The listing defines a tool function \tsf{createTimer} with three parameters---\tsf{name}, \tsf{offset}, and \tsf{period}---and specifies the corresponding syntax for a timer in Lingua Franca.
Both the function's purpose and the parameter descriptions are provided in natural language.
When following this approach, a separate function must be provided for all syntactic constructs that should be exposed via the Tool \ac{api} to the \ac{llm}.
Which granularity works best is an open question and may depend on the syntax of the \ac{dsl}.

For the prototype, the Tool \ac{api} calls have been created manually, but the description of the Tool \ac{api} and the \acp{dsl} grammar already show similarities that can be exploited. 
\autoref{lst:lf-timer-grammar} shows the definition of the timer in the current Lingua Franca grammar\footnote{https://github.com/lf-lang/lingua-franca/blob/master/core/src/main/java/org/lflang/\\LinguaFranca.xtext}.
From this, one can derive that without attributes there are three parameters and the example explains their usage.
The Tool \ac{api} definition can be generated automatically from this grammar definition.
We therefore encourage tool developers to adopt a formal specification that allows for explicitly associating comments with their respective parameters, enabling seamless automation of this process.

Even without a formal definition, portions of the grammar can already be leveraged to generate Tool \ac{api} definitions based solely on existing comments.
For the timer in \autoref{lst:lf-timer-grammar}, ChatGPT\footnote{https://chatgpt.com} generates the result shown in \autoref{lst:lf-timer-tool-api-generated}, which closely resembles the manually written description in \autoref{lst:lf-timer-tool-api}.

\section{Related Work}
\label{sec:related-work}

Tian et al.~\cite{TianLLT+23} investigate ChatGPT's capabilities in code generation, program repair, and code summarization.
Their findings indicate that while ChatGPT struggles with novel or unseen problems—particularly when using long prompts—it performs competitively compared to state-of-the-art program repair tools.
In a related context, Tihanyi et al.~\cite{TihanyiJCF+23} propose a continuous improvement cycle in which code is inspected, model-checked, and counterexamples are fed back into the \ac{llm}.
Both approaches resonate with our emphasis on iterative feedback and refinement.

Sobania et al.~\cite{SobaniaBHP23} argue that AI-based programming, such as with ChatGPT, is currently only effective when paired with a human developer—a view that aligns with our interactive design philosophy.

Our approach is also conceptually related to the chain-of-thought prompting method introduced by Wei et al.\cite{WeiWSM+22}, although from a different perspective.
Whereas chain-of-thought operates internally within \acp{llm} to improve reasoning, our work focuses on external, modeler-visible reasoning across transformation stages.
This is similar in spirit to model-based compiler technology, which aims to make compilation steps transparent and observable~\cite{SmythSRvH18b}.

KIELER SCCharts~\cite{vonHanxledenDM+14}, previously introduced in \autoref{sec:introduction}, is a Statecharts dialect designed for safety-critical systems and uses the same transient view technology as Lingua Franca.
The concept described in \autoref{sec:future-of-modeling} should be directly applicable to SCCharts.
Due to the modular nature of our approach, other domain-specific languages may also adopt selected components---such as automatic diagram synthesis or interaction points---incrementally.

Recent advancements in \ac{ai} workflows---such as specification-driven development exemplified by Kiro\footnote{https://kiro.dev}---emphasize plan formulation prior to execution. These trend represents an orthogonal direction to our approach. Any technique that enhances the output quality of the \ac{llm} can be considered complementary. Determining the optimal combination of methods w.r.t. fine-grained user interaction remains an open question.

\section{Conclusion}
\label{sec:conclusion}

The current state of AI-assisted programming in general-purpose languages increasingly empowers developers to iteratively refine code and inspect changes—a workflow exemplified by tools such as Cursor, where users engage in dialog-driven interactions with \acp{llm}.
In this work, we explored how similar refinement workflows can be extended to the development of domain-specific languages.
Textual and graphical \ac{dsl}-oriented \acp{ide} can both benefit from AI assistance, especially when combined with visual representations that support intuitive inspection~\cite{SmythSRvH19}, seamless navigation, and new \emph{interaction points}, such as voice-based dialogs.

We addressed key challenges in \ac{ai}-assisted \ac{dsl} modeling, including the limitations of one-shot prompting for complex languages and the difficulty of generating graphical \acp{dsl}.
To mitigate these issues, we proposed a toolchain that introduces additional observation and interaction points and delegates contextualization to the Tool \ac{api}.
This enables fine-grained, stage-wise program/model refinement in place of relying solely on extended prompts.
This approach is orthogonal to further improvements of the \acp{llm} themselves through fine-tuning and other retrieval-based techniques. 

We demonstrated this approach with a working prototype implemented as a Visual Studio Code extension for the Lingua Franca language.
By integrating \acp{llm} with voice transcription, reasoning capabilities, and automatic diagram synthesis, our prototype enables a more interactive and reliable modeling experience.
The approach is compatible with modern \acp{ide} and paves the way for future research into more adaptive and intelligent modeling environments.

\bibliographystyle{ACM-Reference-Format}
{
	\bibliography{bib/lf-nli-paper}


\begin{thebibliography}{13}


\ifx \showCODEN    \undefined \def \showCODEN     #1{\unskip}     \fi
\ifx \showISBNx    \undefined \def \showISBNx     #1{\unskip}     \fi
\ifx \showISBNxiii \undefined \def \showISBNxiii  #1{\unskip}     \fi
\ifx \showISSN     \undefined \def \showISSN      #1{\unskip}     \fi
\ifx \showLCCN     \undefined \def \showLCCN      #1{\unskip}     \fi
\ifx \shownote     \undefined \def \shownote      #1{#1}          \fi
\ifx \showarticletitle \undefined \def \showarticletitle #1{#1}   \fi
\ifx \showURL      \undefined \def \showURL       {\relax}        \fi
\providecommand\bibfield[2]{#2}
\providecommand\bibinfo[2]{#2}
\providecommand\natexlab[1]{#1}
\providecommand\showeprint[2][]{arXiv:#2}

\bibitem[Abrial(2005)]%
        {Abrial05}
\bibfield{author}{\bibinfo{person}{Jean-Raymond Abrial}.} \bibinfo{year}{2005}\natexlab{}.
\newblock \showarticletitle{Steam-boiler control specification problem}.
\newblock \bibinfo{journal}{\emph{Formal Methods for Industrial Applications: Specifying and Programming the Steam Boiler Control}} (\bibinfo{year}{2005}), \bibinfo{pages}{500--509}.
\newblock


\bibitem[Busch et~al\mbox{.}(2025)]%
        {BuschBSS24}
\bibfield{author}{\bibinfo{person}{Daniel Busch}, \bibinfo{person}{Alexander Bainczyk}, \bibinfo{person}{Steven Smyth}, {and} \bibinfo{person}{Bernhard Steffen}.} \bibinfo{year}{2025}\natexlab{}.
\newblock \showarticletitle{LLM-Based Code Generation and System Migration in Language-Driven Engineering}. In \bibinfo{booktitle}{\emph{STTT}}. Springer, \bibinfo{pages}{375--390}.
\newblock


\bibitem[Busch et~al\mbox{.}(2023)]%
        {BuschNBS23}
\bibfield{author}{\bibinfo{person}{Daniel Busch}, \bibinfo{person}{Gerrit Nolte}, \bibinfo{person}{Alexander Bainczyk}, {and} \bibinfo{person}{Bernhard Steffen}.} \bibinfo{year}{2023}\natexlab{}.
\newblock \showarticletitle{ChatGPT in the loop: a natural language extension for domain-specific modeling languages}. In \bibinfo{booktitle}{\emph{International Conference on Bridging the Gap between AI and Reality}}. Springer, \bibinfo{pages}{375--390}.
\newblock


\bibitem[Lohstroh et~al\mbox{.}(2021)]%
        {LohstrohMBL21}
\bibfield{author}{\bibinfo{person}{Marten Lohstroh}, \bibinfo{person}{Christian Menard}, \bibinfo{person}{Soroush Bateni}, {and} \bibinfo{person}{Edward~A Lee}.} \bibinfo{year}{2021}\natexlab{}.
\newblock \showarticletitle{Toward a lingua franca for deterministic concurrent systems}.
\newblock \bibinfo{journal}{\emph{ACM Transactions on Embedded Computing Systems (TECS)}} \bibinfo{volume}{20}, \bibinfo{number}{4} (\bibinfo{year}{2021}), \bibinfo{pages}{1--27}.
\newblock


\bibitem[Schneider et~al\mbox{.}(2013)]%
        {SchneiderSvH13}
\bibfield{author}{\bibinfo{person}{Christian Schneider}, \bibinfo{person}{Miro Sp{\"o}nemann}, {and} \bibinfo{person}{Reinhard von Hanxleden}.} \bibinfo{year}{2013}\natexlab{}.
\newblock \showarticletitle{Just Model! -- {P}utting Automatic Synthesis of Node-Link-Diagrams into Practice}. In \bibinfo{booktitle}{\emph{Proceedings of the IEEE Symposium on Visual Languages and Human-Centric Computing (VL/HCC '13)}}. \bibinfo{publisher}{IEEE}, \bibinfo{address}{San Jose, CA, USA}, \bibinfo{pages}{75--82}.
\newblock
\showISSN{1943-6092}
\href{https://doi.org/10.1109/VLHCC.2013.6645246}{doi:\nolinkurl{10.1109/VLHCC.2013.6645246}}


\bibitem[Smyth et~al\mbox{.}(2018)]%
        {SmythSRvH18b}
\bibfield{author}{\bibinfo{person}{Steven Smyth}, \bibinfo{person}{Alexander Schulz-Rosengarten}, {and} \bibinfo{person}{Reinhard von Hanxleden}.} \bibinfo{year}{2018}\natexlab{}.
\newblock \showarticletitle{Towards Interactive Compilation Models}. In \bibinfo{booktitle}{\emph{Proceedings of the 8th International Symposium on Leveraging Applications of Formal Methods, Verification and Validation (ISoLA 2018)}} \emph{(\bibinfo{series}{LNCS}, Vol.~\bibinfo{volume}{11244})}. \bibinfo{publisher}{Springer}, \bibinfo{address}{Limassol, Cyprus}, \bibinfo{pages}{246--260}.
\newblock
\href{https://doi.org/10.14279/tuj.eceasst.78.1098}{doi:\nolinkurl{10.14279/tuj.eceasst.78.1098}}


\bibitem[Smyth et~al\mbox{.}(2019)]%
        {SmythSRvH19}
\bibfield{author}{\bibinfo{person}{Steven Smyth}, \bibinfo{person}{Alexander Schulz-Rosengarten}, {and} \bibinfo{person}{Reinhard von Hanxleden}.} \bibinfo{year}{2019}\natexlab{}.
\newblock \showarticletitle{Guidance in model-based compilations}.
\newblock \bibinfo{journal}{\emph{Electronic Communications of the EASST}}  \bibinfo{volume}{78} (\bibinfo{year}{2019}).
\newblock


\bibitem[Sobania et~al\mbox{.}(2023)]%
        {SobaniaBHP23}
\bibfield{author}{\bibinfo{person}{Dominik Sobania}, \bibinfo{person}{Martin Briesch}, \bibinfo{person}{Carol Hanna}, {and} \bibinfo{person}{Justyna Petke}.} \bibinfo{year}{2023}\natexlab{}.
\newblock \showarticletitle{An analysis of the automatic bug fixing performance of chatgpt}. In \bibinfo{booktitle}{\emph{2023 IEEE/ACM International Workshop on Automated Program Repair (APR)}}. IEEE, \bibinfo{pages}{23--30}.
\newblock


\bibitem[Steffen et~al\mbox{.}(2011)]%
        {SteffenHM11}
\bibfield{author}{\bibinfo{person}{Bernhard Steffen}, \bibinfo{person}{Falk Howar}, {and} \bibinfo{person}{Maik Merten}.} \bibinfo{year}{2011}\natexlab{}.
\newblock \showarticletitle{Introduction to active automata learning from a practical perspective}.
\newblock \bibinfo{journal}{\emph{Formal Methods for Eternal Networked Software Systems: 11th International School on Formal Methods for the Design of Computer, Communication and Software Systems, SFM 2011, Bertinoro, Italy, June 13-18, 2011. Advanced Lectures 11}} (\bibinfo{year}{2011}), \bibinfo{pages}{256--296}.
\newblock


\bibitem[Tian et~al\mbox{.}(2023)]%
        {TianLLT+23}
\bibfield{author}{\bibinfo{person}{Haoye Tian}, \bibinfo{person}{Weiqi Lu}, \bibinfo{person}{Tsz~On Li}, \bibinfo{person}{Xunzhu Tang}, \bibinfo{person}{Shing-Chi Cheung}, \bibinfo{person}{Jacques Klein}, {and} \bibinfo{person}{Tegawend{\'e}~F Bissyand{\'e}}.} \bibinfo{year}{2023}\natexlab{}.
\newblock \showarticletitle{Is ChatGPT the ultimate programming assistant--how far is it?}
\newblock \bibinfo{journal}{\emph{arXiv preprint arXiv:2304.11938}} (\bibinfo{year}{2023}).
\newblock


\bibitem[Tihanyi et~al\mbox{.}(2023)]%
        {TihanyiJCF+23}
\bibfield{author}{\bibinfo{person}{Norbert Tihanyi}, \bibinfo{person}{Ridhi Jain}, \bibinfo{person}{Yiannis Charalambous}, \bibinfo{person}{Mohamed~Amine Ferrag}, \bibinfo{person}{Youcheng Sun}, {and} \bibinfo{person}{Lucas~C Cordeiro}.} \bibinfo{year}{2023}\natexlab{}.
\newblock \showarticletitle{A new era in software security: Towards self-healing software via large language models and formal verification}.
\newblock \bibinfo{journal}{\emph{arXiv preprint arXiv:2305.14752}} (\bibinfo{year}{2023}).
\newblock


\bibitem[von Hanxleden et~al\mbox{.}(2014)]%
        {vonHanxledenDM+14}
\bibfield{author}{\bibinfo{person}{Reinhard von Hanxleden}, \bibinfo{person}{Bj{\"o}rn Duderstadt}, \bibinfo{person}{Christian Motika}, \bibinfo{person}{Steven Smyth}, \bibinfo{person}{Michael Mendler}, \bibinfo{person}{Joaqu{\'i}n Aguado}, \bibinfo{person}{Stephen Mercer}, {and} \bibinfo{person}{Owen O'Brien}.} \bibinfo{year}{2014}\natexlab{}.
\newblock \showarticletitle{{SCCharts: Sequentially Constructive Statecharts} for Safety-Critical Applications}. In \bibinfo{booktitle}{\emph{Proc.\ ACM SIGPLAN Conference on Programming Language Design and Implementation (PLDI '14)}}. \bibinfo{publisher}{ACM}, \bibinfo{address}{Edinburgh, UK}, \bibinfo{pages}{372--383}.
\newblock


\bibitem[Wei et~al\mbox{.}(2022)]%
        {WeiWSM+22}
\bibfield{author}{\bibinfo{person}{Jason Wei}, \bibinfo{person}{Xuezhi Wang}, \bibinfo{person}{Dale Schuurmans}, \bibinfo{person}{Maarten Bosma}, \bibinfo{person}{Fei Xia}, \bibinfo{person}{Ed Chi}, \bibinfo{person}{Quoc~V Le}, \bibinfo{person}{Denny Zhou}, {et~al\mbox{.}}} \bibinfo{year}{2022}\natexlab{}.
\newblock \showarticletitle{Chain-of-thought prompting elicits reasoning in large language models}.
\newblock \bibinfo{journal}{\emph{Advances in neural information processing systems}}  \bibinfo{volume}{35} (\bibinfo{year}{2022}), \bibinfo{pages}{24824--24837}.
\newblock


\end{thebibliography}
}

\end{document}